\documentclass{bmvc2k}

\usepackage{microtype}
\usepackage{graphicx}
\usepackage{subfigure}
\usepackage{booktabs} 
\usepackage{wrapfig}
\usepackage{hyperref}
\usepackage{multirow}
\usepackage{tabularray}
\usepackage{adjustbox}
\usepackage{bm}

\usepackage[ruled,linesnumbered]{algorithm2e}

\usepackage{amsmath}
\usepackage{amssymb}
\usepackage{mathtools}
\usepackage{amsthm}

\usepackage[capitalize,noabbrev]{cleveref}

\title{TACTFL: Temporal Contrastive Training for Multi-modal Federated Learning with Similarity-guided Model Aggregation}

\addauthor{Guanxiong Sun}{guanxiong.sun@bristol.ac.uk}{1}
\addauthor{Majid Mirmehdi}{m.mirmehdi@bristol.ac.uk}{2}
\addauthor{Zahraa Abdallah}{zahraa.abdallah@bristol.ac.uk}{1}
\addauthor{Raul Santos-Rodriguez}{enrsr@bristol.ac.uk}{1}
\addauthor{Ian Craddock}{ian.craddock@bristol.ac.uk}{1,3}
\addauthor{Telmo de Menezes e Silva Filho}{telmo.silvafilho@bristol.ac.uk}{1}

\addinstitution{
 School of Engineering Mathematics and Technology\\
 University of Bristol\\
 Bristol, UK
}
\addinstitution{
 School of Computer Science\\
 University of Bristol\\
 Bristol, UK
}
\addinstitution{
 School of Civil, Aerospace and Design Engineering\\
 University of Bristol\\
 Bristol, UK
}

\runninghead{Student, Prof, Collaborator}{BMVC Author Guidelines}

\begin{document}

\maketitle

\begin{abstract}

Real-world federated learning faces two key challenges: limited access to labelled data and the presence of heterogeneous multi-modal inputs. This paper proposes TACTFL, a unified framework for semi-supervised multi-modal federated learning. TACTFL introduces a modality-agnostic temporal contrastive training scheme that conducts representation learning from unlabelled client data by leveraging temporal alignment across modalities. However, as clients perform self-supervised training on heterogeneous data, local models may diverge semantically. To mitigate this, TACTFL incorporates a similarity-guided model aggregation strategy that dynamically weights client models based on their representational consistency, promoting global alignment. Extensive experiments across diverse benchmarks and modalities, including video, audio, and wearable sensors, demonstrate that TACTFL achieves state-of-the-art performance. For instance, on the UCF101 dataset with only 10\% labelled data, TACTFL attains 68.48\% top-1 accuracy, significantly outperforming the FedOpt baseline of 35.35\%. Code will be released upon publication.

\end{abstract}

\section{Introduction}
\label{sec:intro}

Federated learning (FL)~\cite{fedavg} is a decentralised machine learning paradigm for training models on client devices collaboratively, without transferring clients' raw data to a central server. This privacy-preserving setting has become increasingly important in data-sensitive applications such as healthcare~\cite{health1,health2,health3}, internet of things~\cite{iot1,iot2}, and finance~\cite{finance1,finance2}. Traditional FL methods typically rely on two restrictive assumptions: (i) well-annotated client data, and (ii) single-modality input data. These assumptions often fail in practice due to annotation costs, privacy constraints, and diverse multi-modal data generation by modern devices (e.g., smartphones, wearables, smart cameras). Thus, recent studies have explored semi-supervised~\cite{semifl,semipfl}, unsupervised~\cite{flds,mmproto}, as well as multi-modal FL (MMFL)~\cite{mmfed,fedmm}.
In this paper, we investigate a practically important yet challenging scenario: \textit{semi-supervised MMFL}, where clients hold unlabelled multi-modal data, and a limited labelled dataset is available only on the server. Prior work has explored this scenario~\cite{mmfed,creamfl,ssfl4ar,mfcpl}, but significant challenges remain unresolved, specifically, learning effective multi-modal representations from unlabelled data and handling semantic misalignment in model aggregation.

\begin{figure*}[t]
\begin{center}
\centerline{\includegraphics[width=\linewidth]{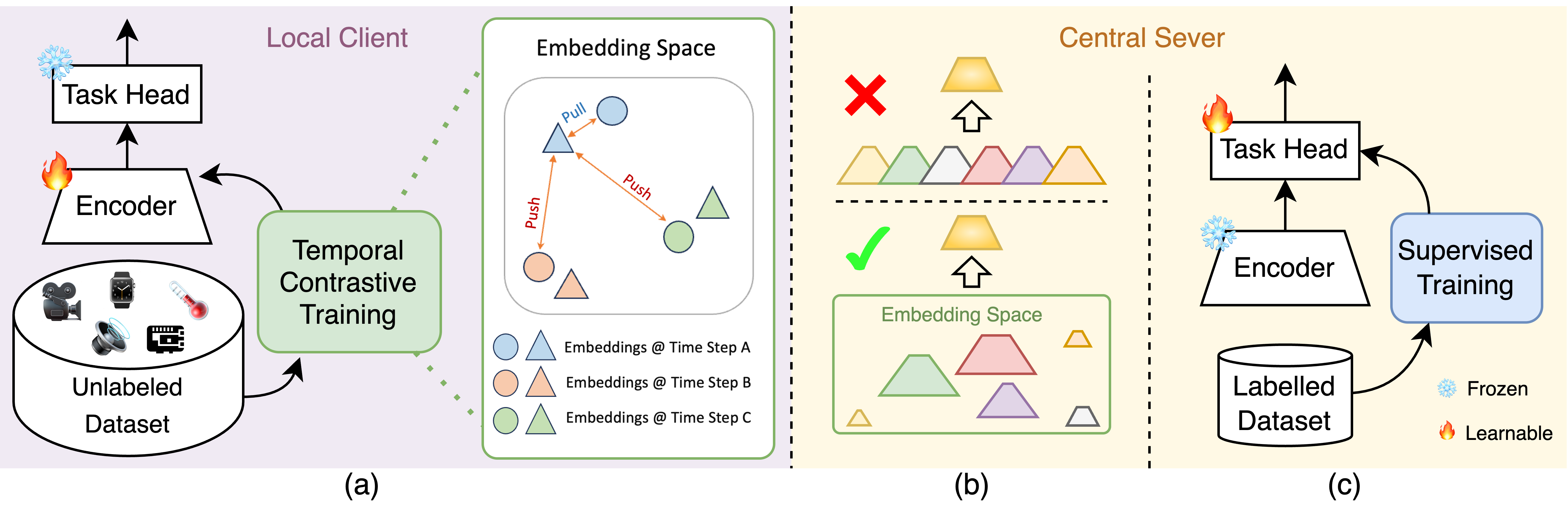}}
\caption{A conceptual overview of TACTFL. (a) On each client, encoders are updated using temporal contrastive training on unlabelled multi-modal data, while task heads remain frozen. The objective encourages embeddings from different modalities at the same time step to be pulled together in the feature space, and embeddings from different time steps to be pushed apart. (b) Unlike FedAvg~\cite{fedavg}, TACTFL adaptively weights each client model based on representational similarity. (c) The server trains the task head on a small labelled dataset with frozen encoders.}
\label{fig:concept}           
\end{center}
\end{figure*}

To address these challenges, we propose TACTFL: TemporAl Contrastive Training for multi-modal Federated Learning, a unified framework that consists of two key components. The \textit{first component} is a novel local temporal contrastive training strategy. Many semi-supervised MMFL methods ~\cite{ssfl4ar,mmfed,optmmfed,fedmekt,fedrecon} rely on reconstruction-based training, which is difficult to optimise and can lack semantic relevance~\cite{ae1,ae2}. Therefore, recent methods explore contrastive learning~\cite{simclr,moco,mocov2} and achieve promising results.
However, CreamFL~\cite{creamfl} works on image-text pairs and requires expensive cross-client intermediate representations exchanges. Similarly, MFCPL~\cite{mfcpl} requires class labels to construct class-specific prototypes and introduces extra communication costs when exchanging class prototypes across clients.
FUELS~\cite{fuels} is tailored for spatio-temporal forecasting tasks in the domain of traffic prediction, and thus, the input data should be well-structured, unimodal data.
FTCCL~\cite{fedtccl} is designed for unimodal time series data, e.g., accelerometer signal, and it particularly focuses on the task of fault diagnosis.
In contrast, TACTFL supports multi-modal data (e.g., video-audio streams or wearable sensor signals) and introduces a modal-agnostic supervision signal via temporal intersection-over-union (tIoU). Specifically, positive pairs are softly weighted based on their temporal overlap, providing richer, more flexible guidance for representation alignment across modalities. 
This departs from existing contrastive learning based MMFL approaches that rely on specific modalities~\cite{fedtccl,fuels} and hard-labelled positives~\cite{creamfl,mfcpl}, enabling more flexible, graded alignment across modalities. To our knowledge, this is the first use of tIoU as a soft contrastive signal for MMFL, making our framework modality-agnostic.
The conceptual overview of local temporal contrastive training is shown in~\cref{fig:concept}(a). After local training, local models are sent to the central server without introducing any extra communication costs (intermediate representations or prototypes).

Our \textit{second component} is then a similarity-guided model aggregation process to mitigate semantic misalignment between client models, an issue largely unaddressed in existing MMFL literature, which typically employs standard aggregation methods (e.g., FedAvg~\cite{fedavg}, FedOpt~\cite{fedopt}).
In TACTFL, the server adaptively weights each client’s model contribution based on representational similarity. Specifically, we compute pairwise cosine similarities between client model representations obtained using the small labelled dataset on the server. Clients that produce representations more consistent with the majority (i.e., higher inter-client similarity) are considered more semantically aligned and are given greater weight during aggregation. The comparison between the standard aggregation technique and the similarity-guided model aggregation is shown as ~\cref{fig:concept}(b). Finally, the central sever updates the task head by supervised training on the small labelled dataset, as shown in ~\cref{fig:concept}(c).

The main contributions of TACTFL are as follows. First, we propose a novel temporal contrastive training scheme, which is modality-agnostic, explicitly leverages temporal alignment across modalities, and avoids additional intermediate representation exchanges. Second, we introduce similarity-guided model aggregation, a dynamic server-side strategy that adaptively weights client models based on representation similarity, effectively addressing semantic misalignment in heterogeneous data scenarios. 
Finally, we conduct extensive experiments across various multi-modal benchmarks spanning video, audio, and wearable sensor data, demonstrating that TACTFL consistently achieves state-of-the-art performance.

\section{Related Work}
\label{sec:rw}

Our work falls under the category of \textit{semi-supervised, multi-modal federated learning}.
In this section, we review relevant literature on FL with a particular focus on semi-supervised learning, MMFL, and contrastive learning approaches. Traditional FL studies mainly explore the challenge  of data heterogeneity~\cite{rethink_data,heter_fl,heter_fl2,tackl_heter}, and aggregation strategies~\cite{fedavg,fedopt,fedprox}. FedAvg~\cite{fedavg} is the earliest FL algorithm. It aggregates local models by computing a weighted average, where each model's contribution is weighted by the size of its local dataset. FedOpt~\cite{fedopt} extends FedAvg by incorporating adaptive optimisation techniques, i.e., momentum, during aggregation. In many real-world settings, local data usually lacks labels, so recent efforts have explored \textit{semi-supervised}~\cite{semidfl,semifed,semifl,semipfl} and \textit{unsupervised}~\cite{lufl,mmproto} FL paradigms, where only a subset of data is labelled or none at all. Survey papers~\cite{fed_survey,semi_survey,semi_survey2,mm_survey} offer comprehensive overviews and trends in FL research.

MMFL extends FL to settings where data from multiple modalities -- such as video, audio, and wearable sensors -- are distributed across clients. FedMEKT~\cite{fedmekt},  FedMAC~\cite{fedmac}, and earlier MMFL approaches~\cite{mmfed,ssfl4ar,optmmfed} adopt deep autoencoder-based objectives to align or reconstruct cross-modal representations. However, reconstruction-based objectives are difficult to optimise~\cite{ae1,ae2} and often misaligned with downstream tasks. Similarly, FedRecon~\cite{fedrecon} focuses on reconstructing missing modalities but assumes strong assumptions on modality mappings. MFCPL~\cite{mfcpl} addresses missing modalities and data heterogeneity via a cross-prototype learning framework, but operates under a supervised learning paradigm, requiring class labels to construct class-specific prototypes.

Contrastive learning has become popular for representation learning from unlabelled data~\cite{simclr,moco}. Recent MMFL methods adopt contrastive objectives to overcome the limitations of autoencoders. CreamFL~\cite{creamfl} aligns image-text pairs through contrastive learning but requires sharing intermediate representations between clients and server, increasing communication overhead. MOON~\cite{moon} and FedMAC~\cite{fedmac} use model-contrastive objectives to align local and global representations, but they are designed for unimodal settings.
More recent works, such as FUELS~\cite{fuels}, target personalised federated spatio-temporal forecasting via dual semantic alignment but require both modality-specific encoders and global prototypes. Similarly, FTCCL~\cite{fedtccl} introduces temporal-context contrastive learning for fault diagnosis but focuses on supervised or pseudo-label training and does not consider modality alignment or missing modalities.

In contrast to prior works, by conducting temporal contrastive training, TACTFL combines the strengths of contrastive learning with a semi-supervised, modality-agnostic design. TACTFL avoids any intermediate representation or prototype exchange, which preserves privacy and minimises communication overhead. Moreover, TACTFL accommodates missing modalities during training and inference and can remain robust.

\section{Method}

TACTFL follows an iterative training workflow. First, the server initialises the global encoders and task head and distributes them to all clients. Each client then performs local encoder training using unlabelled multi-modal data with temporal contrastive training, keeping the task head frozen, as shown in~\cref{fig:concept}(a). After local training, clients upload their updated encoder weights to the server, which aggregates them using the similarity-guided model aggregation strategy, as shown in~\cref{fig:concept}(b). The server then trains the task head using the small labelled dataset while keeping the encoders fixed, as shown in~\cref{fig:concept}(c). Finally, the updated model (encoders and task head) is redistributed to all clients. This process is repeated multiple rounds. 
A detailed algorithm can be found in the Supplementary Material.

\subsection{Problem Definition}
\label{subsec:pre}

We consider a federated learning scenario with $C$ clients, each holding locally collected multi-modal data from heterogeneous sources such as smart cameras, wearable sensors, and mobile devices. Each client is assumed to possess at least two synchronised modalities, denoted as $x^A$ and $x^B$. These may include RGB video frames (modality A), audio signals (modality B), or time-series data such as accelerometer and gyroscope measurements. In practice, some clients may experience missing modalities.

Let $\mathcal{D}_c = \{(x_i^A, x_i^B)\}_{i=1}^{N_c}$ represent the local unlabelled dataset on client $c \in \{1, \dots, C\}$. All client datasets $\mathcal{D}_1, \dots, \mathcal{D}_C$ are fully unlabelled, reflecting real-world challenges in data annotation at the edge. In contrast, a small labelled proxy dataset is available on the central server: $\mathcal{D}_S = \{(x_j^A, x_j^B, y_j)\}_{j=1}^{N_s}$, where $y_j$ denotes task-specific labels. This server-side dataset shares the same modality structure as the client data and serves two key purposes: (i) training the downstream task head, and (ii) supporting similarity-based model aggregation to address inter-client semantic misalignment.
The objective is to collaboratively learn a global model that can perform a downstream task (e.g., action recognition or emotion classification) by exploiting unlabelled multi-modal data from clients and the small labelled dataset on the server, without transferring raw data. This setup ensures strong privacy preservation while enabling effective and communication-efficient training.

\begin{figure*}[t]
\begin{center}
\centerline{\includegraphics[width=\linewidth]{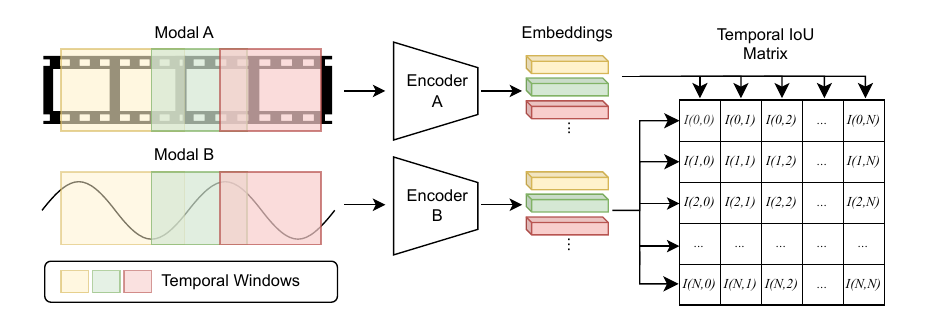}}
\caption{Details of the temporal contrastive training process. }
\label{fig:tct}           
\end{center}
\end{figure*}

\textbf{Generalised Model Architecture.} TACTFL adopts a modular and lightweight architecture composed of three parts: modality-specific encoders, a fusion module, and a downstream task head.
For each input modality, an encoder $E^m$ (e.g., $E^A$ for RGB, $E^B$ for audio), as shown in ~\cref{fig:tct}, is used to extract high-level representations from raw data. The encoders can be implemented using modality and application-appropriate architectures, such as CNNs for visual data or LSTMs for wearable signals, and are trained locally on unlabelled data using our temporal contrastive training strategy, while the task head remains fixed during this stage.

The outputs of the encoders are aggregated by a lightweight fusion module that combines per-modality embeddings into a unified representation. This fusion can be implemented via simple concatenation or averaging, though more advanced approaches (e.g., attention-based fusion) are also compatible.
The fused embedding is then passed to a prediction head $H$, which is responsible for producing task-related outputs. Unlike the encoders, the task head is trained exclusively on the central server using a small labelled proxy dataset. During this supervised training phase, encoder weights are frozen to preserve the representations learned during local contrastive training.

\subsection{Temporal Contrastive Training}
\label{subsec:contrastive}

To enable local learning on unlabelled multi-modal data, we introduce a self-supervised contrastive objective that leverages temporal alignment as implicit supervision. The idea is that signals captured from different modalities within overlapping time windows are likely to describe the same underlying activity and should be embedded closely in the embedding space. ~\cref{fig:concept}(a) shows some examples in the embedding space.

The temporal contrastive training process is shown as~\cref{fig:tct}. For each multi-modal sample, we segment the input sequence into two temporal chunks. The window size is defined as a proportion of the full input (e.g., 80\% or 90\%), and segments may overlap to ensure smooth transitions. Each segment preserves temporal coherence and can be treated as an individual training sample. For a client with two modalities (e.g., RGB video and audio), we denote the resulting sets of temporal segments as $X^A = \{x^A_k\}_{k=0}^{2N-1}$ and $X^B = \{x^B_k\}_{k=0}^{2N-1}$, where $2N$ is the total number of segments across the batch. For each pair $(x^A_i, x^B_j)$, we compute their tIoU, a widely used metric in action localisation. The tIoU serves as a soft pseudo-label for the likelihood that the two segments represent the same content.
We construct a matrix of tIoU scores $I(i, j)$ between all segment pairs from the two modalities, and normalise each row to produce a probability distribution. $T(i, \cdot)$ represents a soft target distribution for $x^A_i$ across all $x^B_j$, and $T(i, j)$ is computed as ${I(i, j)}/{\sum_{k=0}^{2N} I(i, k)}$.
The contrastive loss encourages the embeddings of segments with high temporal overlap to be closer in the feature space, pushing apart unrelated ones. The loss for a pair $(i, j)$ is defined as
\begin{equation}
    l(i, j) = -\log \frac{e^{\mathcal{S}(i, j) / \tau}}{\sum_{k=0}^{2N} e^{\mathcal{S}(i, k) / \tau}} T(i, j),
\end{equation}
where $\tau$ is a temperature parameter and $\mathcal{S}(i, j)$ denotes the cosine similarity between encoder outputs as
\begin{equation}
\mathcal{S}(i, j) = \frac{E^A(x^A_i)^\top E^B(x^B_j)}{\|E^A(x^A_i)\| \cdot \|E^B(x^B_j)\|}.
\end{equation}
The total contrastive loss is computed symmetrically over both modalities, aggregating all pairs $(i, j)$ and $(j, i)$ within each mini-batch.
This formulation enables robust cross-modal representation learning without requiring labels, and naturally supports clients with missing modalities: when only a single modality is available, the same modality can be duplicated to form pseudo-pairs for training. Additionally, the temporal segmentation process implicitly augments data and improves generalisation across clients with limited local data.

\subsection{Similarity-guided Model Aggregation}
\label{subsec:sigma}

In contrast to canonical supervised FL, where local models are supervised by the same label distribution and tend to produce semantically aligned feature spaces, the use of self-supervised contrastive learning in TACTFL introduces a new challenge: {semantic misalignment}. Since each client independently optimises its encoders on local data using contrastive objectives, the resulting representations can diverge significantly across clients. For example, the same input sample may produce embeddings with a small cosine similarity when processed by encoders trained on different clients. This inconsistency makes standard aggregation techniques, e.g., FedAvg~\cite{fedavg}, unreliable.

The similarity-guided model aggregation strategy leverages the small labelled dataset on the central server $D_S$ to assess the alignment of each client model. It assigns higher weights to those clients whose representations are more semantically consistent with the rest. Formally, given $C$ client encoder models $\{m_0, m_1, \dots, m_{C-1}\}$, the server first computes feature embeddings for the labelled dataset $D_S$ by passing it through each client model. For each model $m_i$, the embeddings are averaged across all samples in $D_S$ to produce a summary vector $V_i$. This results in $C$ embedding vectors $\{V_0, V_1, \dots, V_{C-1}\}$, one per client.
The server then computes a pairwise similarity matrix $S \in \mathbb{R}^{C \times C}$, where each element $S(i, j)$ reflects the cosine similarity between the representations $V_i$ and $V_j$, computed as $S(i, j) = {V_i \cdot V_j}/{\|V_i\| \cdot \|V_j\|}$.
To determine how well-aligned each client is with the others, we compute a similarity-based weight vector $W \in \mathbb{R}^C$,
\begin{equation}
W(i) = \frac{\sum_{j=0}^{C-1} S(i, j)}{\sum_{i=0}^{k-1} \sum_{j=0}^{k-1} S(k, j)},
\end{equation}
ensuring that $\sum_{i=0}^{C-1} W(i) = 1$. These weights capture the relative alignment of each client’s representations with the rest of the population. Finally, the global encoder model $m'$ is obtained as a weighted sum of the client models as
$m' = \sum_{i=0}^{C-1} W(i) \cdot m_i$. By adaptively emphasising clients with better-aligned representations, the training convergence speed is accelerated. It also mitigates the impact of noisy, biased, poorly trained local models.

After the model aggregation, the server conducts supervised central training in a standard way. The task head \( H \) is trained using the small labelled dataset \( D_S \). During this stage, the encoders are frozen, and only \( H \) is optimised with supervised loss. The updated model containing the trained head is then sent to clients for the next round. More details of the central training can be found in the Supplementary Material.




\section{Experiments}
\label{experiment}

We evaluate TACTFL against state-of-the-art methods on diverse benchmark datasets covering various modalities and tasks. Following FedMEKT~\cite{fedmekt}, we conduct experiments under two scenarios reflecting real-world challenges: \textit{multi-modal} and \textit{missing-modal} clients.

\textbf{Detailed Settings.} Multi-modal encoders extract embeddings from raw input. Following FedMM~\cite{fedmm}, MobileNetV2~\cite{mobilenet}, mel-frequency cepstral coefficients (MFCCs)~\cite{mfcc}, and MobileBERT~\cite{mobilebert} are used as encoders for videos/images, audios, and text, respectively. 
Following FedMEKT~\cite{fedmekt}, 2 LSTM layers~\cite{lstm} are used as the encoder for wearable data. For our experiments, we follow FedMM~\cite{fedmm} to conduct local training for 1 epoch in each of 200 rounds of global-local updates and fix the batch size for all datasets to 16. The number of clients for each dataset can be found in the Supplementary Material.


\begin{table}[t]
\scriptsize
\begin{center}
\resizebox{\columnwidth}{!}
{%
\begin{tabular}{lllllll|lll}
\toprule
\multirow{2}{*}{\textbf{Dataset}} &
  \multirow{2}{*}{\bm{$\alpha$}} &
  \multirow{2}{*}{\textbf{Modalities}} &
  \multirow{2}{*}{\textbf{Metric}} &
  \multicolumn{3}{c}{\bm{$r_l = 0.9$}} &
  \multicolumn{3}{c}{\bm{$r_l=0.8$}} \\ \cmidrule(l){5-10}
        &                    &              &     & \textbf{FedAvg}~\cite{fedavg} & \textbf{FedOpt}~\cite{fedopt} & {\bf TACTFL} & \textbf{FedAvg}~\cite{fedavg} & \textbf{FedOpt}~\cite{fedopt} & \textbf{TACTFL} \\ \midrule
UCF101~\cite{ucf101}  & 0.1                & Audio, Video  & Acc & 10.49  & \underline{35.35}  & \textbf{68.48}  & 40.14  & \underline{44.43}  & \textbf{71.67}  \\
MiT10~\cite{mit}   &                    & Audio, Video  & Acc & 7.48   & \underline{19.86}  & \textbf{48.87}  & 25.77  & \underline{36.90}  & \textbf{51.77}  \\
UCI-HAR~\cite{ucihar} &                    & Acce, Gyro     & F1  & 33.78  & \underline{41.26}  & \textbf{74.87}  & 35.11  & \underline{42.64}  & \textbf{75.12}  \\ \midrule
UCF101~\cite{ucf101}  & 5.0                & Audio, Video  & Acc & 6.41  & \underline{12.56}  & \textbf{70.22}  & 40.71  & \underline{46.53}  & \textbf{72.99}  \\
MiT10~\cite{mit}   &                    & Audio, Video  & Acc & 9.21   & \underline{21.04}  & \textbf{50.13}  & 27.60  & \underline{38.15}  & \textbf{52.66}  \\
UCI-HAR~\cite{ucihar} &                    & Acce, Gyro     & F1  & 35.66  & \underline{45.21}  & \textbf{79.49}  & 37.70  & \underline{44.50}  & \textbf{76.41}  \\  \bottomrule
\end{tabular}%
}
\end{center}
\caption{Scenario 1: Accuracy/F1 (\%) of TACTFL and two baseline methods, FedAvg/FedOpt, under two missing label rates $r_l$. $\alpha$ denotes the control factor of the Dirichlet distribution for data partitioning (best results in \textbf{bold}, second best \underline{underlined}).}
\label{tab:fedmm_results}
\end{table}

\begin{table}[]
\scriptsize
\begin{center}
\resizebox{\columnwidth}{!}{%
\begin{tabular}{clllllllc}
\toprule
\textbf{Datasets} & \multicolumn{2}{c}{\textbf{Modality}}        & \textbf{FedIoT}~\cite{mmfed} & \textbf{FedProx}~\cite{fedprox} & \textbf{MOON}~\cite{moon}  & \textbf{CreamFL}~\cite{creamfl} & \textbf{FedMEKT}~\cite{fedmekt} & \textbf{TACTFL} \\
\midrule
\multirow{6}{*}{mHealth~\cite{mhealth}} 
& \multirow{2}{*}{Acce-Gyro} & Acce & 63.83 & 67.86 & 64.47 & \underline{68.58} & 68.28 & \textbf{68.90} \\
&                             & Gyro & 63.62 & 64.36 & 64.12 & 64.34 & \underline{64.60} & \textbf{64.75} \\
& \multirow{2}{*}{Acce-Mage} & Acce & 69.99 & 70.00 & 69.71 & 69.51 & \textbf{71.16} & \underline{71.09} \\
&                             & Mage & 68.49 & 69.21 & 68.79 & 69.18 & \underline{71.13} & \textbf{71.84} \\
& \multirow{2}{*}{Gyro-Mage} & Gyro & 65.43 & 65.37 & 65.90 & 66.05 & \textbf{67.10} & \underline{67.21} \\
&                             & Mage & 68.28 & 68.75 & 68.13 & 68.00 & \underline{69.02} & \textbf{69.30} \\
\midrule
\multirow{6}{*}{UR-Fall~\cite{urfall}} 
& \multirow{2}{*}{Acce-RGB}  & Acce  & 61.70 & 65.66 & 65.89 & 66.82 & \textbf{70.66} & \underline{70.43} \\
&                             & RGB   & 57.88 & 59.26 & 62.00 & 62.34 & \underline{66.70} & \textbf{66.92} \\
& \multirow{2}{*}{Acce-Depth} & Acce  & 67.24 & 68.79 & 68.08 & 71.61 & \textbf{72.68} & \underline{72.17} \\
&                             & Depth & 60.76 & 68.08 & 68.16 & \textbf{75.33} & \underline{75.22} & 74.99 \\
& \multirow{2}{*}{RGB-Depth}  & RGB   & 69.88 & 75.60 & 73.38 & \textbf{78.57} & 77.87 & \underline{78.18} \\
&                             & Depth & 67.61 & 70.04 & 66.48 & 69.18 & \underline{70.57} & \textbf{70.86} \\
\midrule
\multirow{2}{*}{Opportunity~\cite{opportunity}} 
& \multirow{2}{*}{Acce-Gyro} & Acce & 71.75 & 72.25 & 72.96 & 73.33 & \textbf{73.50} & \underline{73.37} \\
&                             & Gyro & 72.08 & 71.43 & 72.09 & 72.10 & \underline{72.15} & \textbf{73.12} \\
\bottomrule
\end{tabular}
}
\end{center}
\caption{Scenario 1: F1 scores on FedMEKT~\cite{fedmekt} benchmarks. Paired modalities are for training, and each individual modality is for testing (best in \textbf{bold}, second best \underline{underlined}).}
\label{tab:fediot_results}
\end{table}

\textbf{Scenario 1: Multi-modal Clients.} Each client holds synchronised data from multiple modalities, such as video-audio streams or wearable sensor signals. This setting reflects real-world conditions where multi-modal data is plentiful but labels are unavailable. We first follow FedMM~\cite{fedmm} protocols to prepare datasets including UCF101~\cite{ucf101} and MiT-10~\cite{mit} for video-audio, and UCI-HAR~\cite{ucihar} for accelerometer (Acce) and gyroscope (Gyro) data. To simulate federated non-IID distributions, data is partitioned using a Dirichlet distribution $\text{Dir}(\alpha)$, where lower $\alpha$ values introduce higher heterogeneity. The label missing rate $r_l$ controls the level of label availability. For example, when $r_l$=0.9, 90\% data is treated unlabelled for local training, and 10\% data is randomly sampled as the central labelled set $D_S$.
Table~\ref{tab:fedmm_results} compares TACTFL against FedAvg and FedOpt. TACTFL consistently matches or exceeds baselines.

We further evaluate TACTFL under the FedMEKT~\cite{fedmekt} setup using three human activity datasets. The mHealth dataset~\cite{mhealth} includes accelerometer, gyroscope, and magnetometer (Mage) signals. One participant is held out for testing, one for the labelled set $D_S$, and the rest for training. For UR-Fall~\cite{urfall}, which includes RGB, depth, and accelerometer data, we randomly sample 10\% for testing, 10\% as $D_S$, and leave the remaining for training. The Opportunity dataset~\cite{opportunity} includes Acce and Gyro signals from four participants performing kitchen tasks. Runs 4 and 5 from participants 2 and 3 are used for test, runs 4 and 5 from participants 1 and 4 are used as $D_S$. As shown in~\cref{tab:fediot_results}, TACTFL consistently achieves the best or second-best results across modality pairs, outperforming strong baselines like FedMEKT and CreamFL, especially in sensor-rich datasets like mHealth and UR-Fall where temporal and multi-modal cues are critical.

\textbf{Scenario 2: Missing Modalities.} In practice, some clients may lack access to certain modalities due to hardware limitations or sensor failures. This setting captures the challenge of learning from incomplete multi-modal inputs, with all client data remaining unlabelled.
We evaluate this scenario on three datasets: (1) UCI-HAR~\cite{ucihar}, featuring Acce and Gyro data; (2) Hateful Memes~\cite{hatefulmemes}, a binary classification dataset of 10k image-text pairs; and (3) MELD~\cite{meld}, a multimodal emotion recognition dataset with audio and text. We use four emotion classes (neutral, happy, sad, angry) with the most samples.
Following FedMM~\cite{fedmm} and MFCPL~\cite{mfcpl}, we simulate missing modalities by randomly dropping entire modalities per client. For each client–modality pair, a Bernoulli trial determines whether that modality is retained or replaced with zeros. We experiment with missing modality rates $r_m \in \{0.5, 0.7, 0.8\}$. We report AUC/UAR/F1 metrics for the Hateful Memes/MELD/UCI-HAR datasets, respectively. As shown in Table~\ref{tab:missing_modal_results}, TACTFL achieves the best performance in the majority of cases, and ranks either first or second in every setting. Notably, it outperforms the best state-of-the-art method MFCPL~\cite{mfcpl}, particularly on the MELD dataset. These results highlight TACTFL's robustness and strong generalisation ability in scenarios with missing modalities.

\begin{table}[]
\scriptsize
\begin{center}
\resizebox{\columnwidth}{!}{%
\begin{tabular}{llllllllll}
\toprule
\textbf{Datasets}                       & \bm{$r_m$} & \textbf{FedOpt}~\cite{fedopt} & \textbf{MOON}~\cite{moon}  & \textbf{FedProx}~\cite{fedprox} & \textbf{FPL}~\cite{fpl}   & \textbf{FedIoT}~\cite{mmfed} & \textbf{FedMSplit}~\cite{fedmsplit} & \textbf{MFCPL}~\cite{mfcpl}          & \textbf{TACTFL}         \\
\midrule
\multirow{3}{*}{UCI-HAR~\cite{ucihar}}       & 0.5   & 74.79  & 74.87    & 70.14   & 75.16 & 71.57  & 74.76     & \underline{77.44}          &    \textbf{77.92}            \\
                               & 0.7   & 66.17  & 73.63    & 69.37   & 74.28 & 70.58  & 73.32     & \underline{75.61}          & \textbf{75.83}               \\
                               & 0.8   & 56.45  & 72.78    & 68.77   & 73.46 & 68.62  & 72.77     & \textbf{75.19} & \underline{75.12}    \\ \midrule
\multirow{3}{*}{Hateful Memes~\cite{hatefulmemes}} & 0.5   & 54.62  & 52.67    & 52.75   & 54.97 & 53.98  & 55.57     & \underline{56.30}          &  \textbf{56.89}              \\
                               & 0.7   & 51.61  & 51.40    & 50.49   & 53.79 & 52.72  & 53.61     & \textbf{55.69}         & \underline{55.31}               \\
                               & 0.8   & 49.58  & 50.68    & 50.23   & 52.94 & 51.31  & 52.65     & \underline{55.04}    & \textbf{59.47} \\ \midrule
\multirow{3}{*}{MELD~\cite{meld}}          & 0.5   & 51.58  & 50.66    & 51.92   & 52.11 & 51.81  & 52.97     & \underline{54.71}          &  \textbf{55.26}              \\
                               & 0.7   & 49.27  & 47.51    & 51.83   & 50.09 & 49.48  & 51.65     & \underline{53.62}          &  \textbf{54.10}          \\
                               & 0.8   & 47.54  & 46.94    & 50.48   & 49.49 & 49.61  & 50.57     & \underline{52.18}    & \textbf{53.25} \\ \bottomrule
\end{tabular}%
}
\end{center}
\caption{Scenario 2: Performance for various missing-modality rates ($r_m$) on UCI-HAR, Hateful Memes, MELD datasets (best in \textbf{bold}, second best \underline{underlined}).}
\label{tab:missing_modal_results}
\end{table}


\textbf{Ablation Study.} We perform comprehensive ablation studies on the UCF101 dataset to assess the contribution of each component, including: (1) temporal contrastive training (TCT), (2) similarity-guided model aggregation (SMA), (3) robustness to varying label availability, (4) resilience to missing modalities, and (5) the impact of temporal window size.

\begin{figure}[t]
\begin{center}
\centerline{\includegraphics[width=\columnwidth]{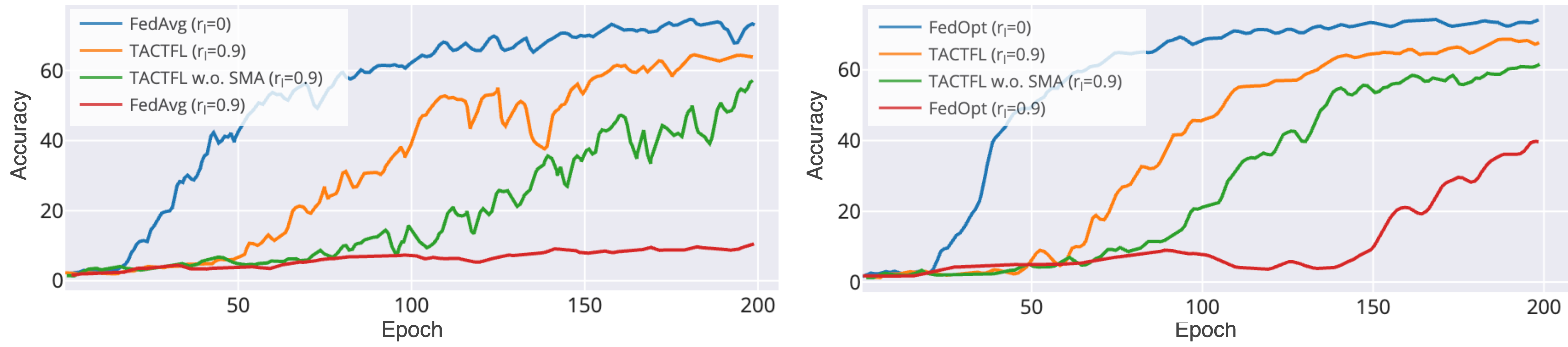}}
\caption{Test accuracy curve under four settings: 1. SSFL ($r_l=0.9$); 2. TACTFL without SMA  ($r_l=0.9$); 3. Full TACTFL ($r_l=0.9$); 4. Fully-supervised FL ($r_l=0$). }
\label{fig:ablation_curve}           
\end{center}
\end{figure}

\textbf{Effects of TCT and SMA.} We evaluate four configurations under both FedAvg and FedOpt settings. First, \textit{SSFL} denotes semi-supervised FL with a missing label rate of $r_l = 0.9$, i.e., only 10\% of the data is labelled and used as $D_S$ on the server, while the remaining 90\% of client data is discarded. Second, \textit{TACTFL w/o SMA} includes TCT on unlabelled data but uses standard aggregation (FedAvg or FedOpt). Lastly, the full \textit{TACTFL} method integrates both TCT and SMA. The test accuracy curves of all configurations are shown in~\cref{fig:ablation_curve}. More specifically, compared to SSFL baselines of FedAvg/FedOpt ($r_l=0.9$), introducing TCT alone (TACTFL w/o SMA) improves accuracy from 10.49\% to 54.58\% and from 40.43\% to 62.91\%, respectively, demonstrating the effectiveness of TCT from unlabelled data. Moreover, incorporating SMA (TACTFL) yields additional performance gains, improving accuracy to 63.12\% (FedAvg) and 69.60\% (FedOpt). This confirms that SMA plays a critical role in mitigating semantic misalignment during aggregation, resulting in a more robust and semantically consistent global model. Furthermore, the test accuracy curve shown in~\cref{fig:ablation_curve} demonstrates that SMA can speed up TACTFL's convergence speed, bringing it close to fully-supervised ($r_l=0$) performance.

\begin{wraptable}{r}{0.5\textwidth}
\begin{center}
\resizebox{\linewidth}{!}{%
\begin{tabular}{llccccc}
\toprule
\multirow{2}{*}{\textbf{Setting}} & \multirow{2}{*}{\textbf{Method}} & \multicolumn{5}{c}{\textbf{Rate}} \\
\cmidrule(l){3-7}
& & \textbf{0.1} & \textbf{0.3} & \textbf{0.5} & \textbf{0.7} & \textbf{0.9} \\
\midrule
\multirow{2}{*}{$r_l$} 
& FedAvg~\cite{fedavg} & 71.02 & 70.81 & 67.28 & 60.25 & 10.49 \\
& TACTFL & 71.31 & 71.15 & 69.73 & 67.08 & 63.12 \\
\midrule
\multirow{2}{*}{$r_m$} 
& FedOpt~\cite{fedopt} & 74.48 & 73.20 & 71.46 & 68.56 & 56.32 \\
& TACTFL & 75.01 & 74.76 & 73.52 & 72.84 & 72.22 \\
\bottomrule
\end{tabular}
}
\end{center}
\caption{Accuracy under varying levels of missing label and missing modality rates, $r_l$ and $r_m$.}
\vskip -0.1in
\label{tab:robustness}
\end{wraptable}

\textbf{Robustness to label and modality sparsity.}
To assess TACTFL's robustness under practical constraints, we evaluate its performance under varying degrees of label scarcity and missing modality scenarios. In  Table~\ref{tab:robustness}, we gradually increase the missing label and missing modality rates, $r_l$ and $r_m$, from 0.1 to 0.9. TACTFL consistently outperforms FedAvg across all values of $r_l$, with the performance gap widening as fewer labels are available. Notably, at $r_l = 0.9$, TACTFL achieves 63.12\% F1 score compared to just 10.49\% for FedAvg. As for missing modalities, even with $r_m=0.9$, TACTFL retains strong performance (72.22\%) and consistently outperforms FedOpt. These results highlight TACTFL’s robustness to both limited supervision and incomplete sensor inputs, making it suitable for deployment in real-world, resource-constrained FL scenarios.

\begin{wraptable}{r}{0.5\textwidth}
\begin{center}
\resizebox{\linewidth}{!}{%
\begin{tabular}{llllll}
\toprule
\textbf{Temporal Window Size} & 90\%  & 80\%  & 70\%  & 60\%  & 50\%  \\
\midrule
\textbf{Accuracy}             & 69.26 & 71.34 & 72.22 & 71.28 & 70.86 \\
\bottomrule
\end{tabular}%
}
\end{center}
\caption{Effect of temporal window size on performance (accuracy \%).}
\label{tab:window_size}
\end{wraptable}

\textbf{Effect of temporal window size.} 
As shown in Table~\ref{tab:window_size}, moderate window sizes (e.g., 70–80\%) yield the highest accuracy, while both shorter and longer windows slightly reduce performance. This suggests that appropriately-sized temporal context is crucial for effective temporal contrastive learning across modalities.

\section{Conclusion}
In this work, we presented TACTFL, a unified framework for semi-supervised multi-modal federated learning. By introducing temporal contrastive training and similarity-guided model aggregation, TACTFL effectively leverages unlabelled multi-modal data while addressing semantic misalignment across clients. Extensive experiments demonstrate that our method achieves consistent and significant improvements over existing baselines across diverse benchmarks, highlighting its potential for real-world privacy-preserving applications. Despite its promising performance, TACTFL has several limitations. First, the current framework assumes access to synchronised multi-modal data, which may not always be feasible in real-world settings. Second, the temporal contrastive training currently relies on fixed window parameters, which might not optimally capture semantic dynamics for all modalities. Future work will explore dynamic temporal alignment techniques, support for asynchronous modalities, and personalised model adaptation for highly heterogeneous client populations. 

\section{Ethical Considerations and Broader Impact}

The benchmark datasets used in our evaluation (e.g., UCF101, Hateful Memes, MELD) may contain inherent demographic and cultural biases, which can lead to models that perform unfairly across different user groups. 
Another potential risk of bias may be introduced by the Similarity-guided Model Aggregation (SMA). SMA could penalise client models trained on data from underrepresented populations, skewing the final global model. While TACTFL's federated design is inherently privacy-preserving, developers must be vigilant about these fairness issues. We strongly advocate for the careful auditing of proxy datasets to ensure they are balanced and representative. Future work should explore fairness-aware and personalised aggregation techniques to mitigate these potential biases before any real-world deployment.

\section{Acknowledgements}
This work was supported by the TORUS Project, EPSRC grant number EP/X036146/1. RSR is funded by the UKRI Turing AI Fellowship EP/V024817/1.

\bibliography{egbib}

\begin{thebibliography}{56}
\providecommand{\natexlab}[1]{#1}
\providecommand{\url}[1]{\texttt{#1}}
\expandafter\ifx\csname urlstyle\endcsname\relax
  \providecommand{\doi}[1]{doi: #1}\else
  \providecommand{\doi}{doi: \begingroup \urlstyle{rm}\Url}\fi

\bibitem[Banos et~al.(2014)Banos, Garcia, Holgado-Terriza, Damas, Pomares, Rojas, Saez, and Villalonga]{mhealth}
Oresti Banos, Rafael Garcia, Juan~A Holgado-Terriza, Miguel Damas, Hector Pomares, Ignacio Rojas, Alejandro Saez, and Claudia Villalonga.
\newblock mhealthdroid: A novel framework for agile development of mobile health applications.
\newblock \emph{Ambient Assisted Living and Daily Activities}, 8868\penalty0 (14):\penalty0 91--98, 2014.

\bibitem[Bao et~al.(2023)Bao, Zhang, Miao, Gong, Hu, Liu, Liu, and Shi]{mmproto}
Guangyin Bao, Qi~Zhang, Duoqian Miao, Zixuan Gong, Liang Hu, Ke~Liu, Yang Liu, and Chongyang Shi.
\newblock Multimodal federated learning with missing modality via prototype mask and contrast.
\newblock \emph{arXiv preprint arXiv:2312.13508}, 2023.

\bibitem[Bulbul et~al.(2018)Bulbul, Cetin, and Dogru]{ucihar}
Erhan Bulbul, Aydin Cetin, and Ibrahim~Alper Dogru.
\newblock Human activity recognition using smartphones.
\newblock In \emph{2018 2nd International Symposium on Multidisciplinary Studies and Innovative Technologies (ISMSIT)}, pages 1--6, 2018.
\newblock \doi{10.1109/ISMSIT.2018.8567275}.

\bibitem[Byrd and Polychroniadou(2020)]{finance2}
David Byrd and Antigoni Polychroniadou.
\newblock Differentially private secure multi-party computation for federated learning in financial applications.
\newblock In \emph{Proceedings of the first ACM international conference on AI in finance}, pages 1--9, 2020.

\bibitem[Chavarriaga et~al.(2013)Chavarriaga, Sagha, Calatroni, Digumarti, Tr{\"o}ster, Mill{\'a}n, and Roggen]{opportunity}
Ricardo Chavarriaga, Hesam Sagha, Alberto Calatroni, Sundara~Tejaswi Digumarti, Gerhard Tr{\"o}ster, Jos{\'e} del~R Mill{\'a}n, and Daniel Roggen.
\newblock The opportunity challenge: A benchmark database for on-body sensor-based activity recognition.
\newblock \emph{Pattern Recognition Letters}, 34\penalty0 (15):\penalty0 2033--2042, 2013.

\bibitem[Chen and Zhang(2022)]{fedmsplit}
Jiayi Chen and Aidong Zhang.
\newblock Fedmsplit: Correlation-adaptive federated multi-task learning across multimodal split networks.
\newblock In \emph{Proceedings of the 28th ACM SIGKDD Conference on Knowledge Discovery and Data Mining}, KDD '22, page 87–96, New York, NY, USA, 2022. Association for Computing Machinery.
\newblock ISBN 9781450393850.
\newblock \doi{10.1145/3534678.3539384}.
\newblock URL \url{https://doi.org/10.1145/3534678.3539384}.

\bibitem[Chen and Rudnicky(2023)]{mfcc}
Li-Wei Chen and Alexander Rudnicky.
\newblock Exploring wav2vec 2.0 fine tuning for improved speech emotion recognition.
\newblock In \emph{ICASSP 2023-2023 IEEE International Conference on Acoustics, Speech and Signal Processing (ICASSP)}, pages 1--5. IEEE, 2023.

\bibitem[Chen and Li(2022)]{optmmfed}
Sijia Chen and Baochun Li.
\newblock Towards optimal multi-modal federated learning on non-iid data with hierarchical gradient blending.
\newblock In \emph{IEEE Conference on Computer Communications (INFOCOM)}, 2022.

\bibitem[Chen et~al.(2020{\natexlab{a}})Chen, Kornblith, Norouzi, and Hinton]{simclr}
Ting Chen, Simon Kornblith, Mohammad Norouzi, and Geoffrey Hinton.
\newblock A simple framework for contrastive learning of visual representations.
\newblock In \emph{International conference on machine learning}, pages 1597--1607. PMLR, 2020{\natexlab{a}}.

\bibitem[Chen et~al.(2020{\natexlab{b}})Chen, Fan, Girshick, and He]{mocov2}
Xinlei Chen, Haoqi Fan, Ross Girshick, and Kaiming He.
\newblock Improved baselines with momentum contrastive learning.
\newblock \emph{arXiv preprint arXiv:2003.04297}, 2020{\natexlab{b}}.

\bibitem[Dai et~al.(2023)Dai, Chen, Li, Heinecke, Sun, and Xu]{tackl_heter}
Yutong Dai, Zeyuan Chen, Junnan Li, Shelby Heinecke, Lichao Sun, and Ran Xu.
\newblock Tackling data heterogeneity in federated learning with class prototypes.
\newblock In \emph{AAAI}, 2023.

\bibitem[Diao et~al.(2022)Diao, Ding, and Tarokh]{semifl}
Enmao Diao, Jie Ding, and Vahid Tarokh.
\newblock Semifl: semi-supervised federated learning for unlabeled clients with alternate training.
\newblock In \emph{NeurIPS}, 2022.

\bibitem[Feng et~al.(2023)Feng, Bose, Zhang, Hebbar, Ramakrishna, Gupta, Zhang, Avestimehr, and Narayanan]{fedmm}
Tiantian Feng, Digbalay Bose, Tuo Zhang, Rajat Hebbar, Anil Ramakrishna, Rahul Gupta, Mi~Zhang, Salman Avestimehr, and Shrikanth Narayanan.
\newblock Fedmultimodal: A benchmark for multimodal federated learning.
\newblock In \emph{ACM KDD}, 2023.

\bibitem[Hao et~al.(2019)Hao, Li, Luo, Xu, Yang, and Liu]{iot2}
Meng Hao, Hongwei Li, Xizhao Luo, Guowen Xu, Haomiao Yang, and Sen Liu.
\newblock Efficient and privacy-enhanced federated learning for industrial artificial intelligence.
\newblock \emph{IEEE Transactions on Industrial Informatics}, 16\penalty0 (10):\penalty0 6532--6542, 2019.

\bibitem[He et~al.(2020)He, Fan, Wu, Xie, and Girshick]{moco}
Kaiming He, Haoqi Fan, Yuxin Wu, Saining Xie, and Ross Girshick.
\newblock Momentum contrast for unsupervised visual representation learning.
\newblock In \emph{CVPR}, 2020.

\bibitem[Hochreiter and Schmidhuber(1997)]{lstm}
Sepp Hochreiter and J{\"u}rgen Schmidhuber.
\newblock Long short-term memory.
\newblock \emph{Neural computation}, 9\penalty0 (8):\penalty0 1735--1780, 1997.

\bibitem[Howard(2017)]{mobilenet}
Andrew~G Howard.
\newblock Mobilenets: Efficient convolutional neural networks for mobile vision applications.
\newblock \emph{arXiv preprint arXiv:1704.04861}, 2017.

\bibitem[Huang et~al.(2022)Huang, Ye, and Du]{heter_fl2}
Wenke Huang, Mang Ye, and Bo~Du.
\newblock Learn from others and be yourself in heterogeneous federated learning.
\newblock In \emph{CVPR}, 2022.

\bibitem[Huang et~al.(2023{\natexlab{a}})Huang, Ye, Shi, Li, and Du]{flds}
Wenke Huang, Mang Ye, Zekun Shi, He~Li, and Bo~Du.
\newblock Rethinking federated learning with domain shift: A prototype view.
\newblock In \emph{2023 IEEE/CVF Conference on Computer Vision and Pattern Recognition (CVPR)}, pages 16312--16322. IEEE, 2023{\natexlab{a}}.

\bibitem[Huang et~al.(2023{\natexlab{b}})Huang, Ye, Shi, Li, and Du]{fpl}
Wenke Huang, Mang Ye, Zekun Shi, He~Li, and Bo~Du.
\newblock Rethinking federated learning with domain shift: A prototype view.
\newblock In \emph{2023 IEEE/CVF Conference on Computer Vision and Pattern Recognition (CVPR)}, pages 16312--16322, 2023{\natexlab{b}}.
\newblock \doi{10.1109/CVPR52729.2023.01565}.

\bibitem[Jin et~al.(2023)Jin, Liu, Chen, and Yang]{semi_survey2}
Yilun Jin, Yang Liu, Kai Chen, and Qiang Yang.
\newblock Federated learning without full labels: A survey.
\newblock \emph{arXiv preprint arXiv:2303.14453}, 2023.

\bibitem[Khurram~Soomro(2012)]{ucf101}
Mubarak~Shah Khurram~Soomro, Amir Roshan~Zamir.
\newblock Ucf101: A dataset of 101 human actions classes from videos in the wild.
\newblock \emph{arXiv preprint arXiv:1212.0402}, 2012.

\bibitem[Kiela et~al.(2020)Kiela, Firooz, Mohan, Goswami, Singh, Ringshia, and Testuggine]{hatefulmemes}
Douwe Kiela, Hamed Firooz, Aravind Mohan, Vedanuj Goswami, Amanpreet Singh, Pratik Ringshia, and Davide Testuggine.
\newblock The hateful memes challenge: Detecting hate speech in multimodal memes.
\newblock \emph{Advances in neural information processing systems}, 33:\penalty0 2611--2624, 2020.

\bibitem[Kwolek and Kepski(2014)]{urfall}
Bogdan Kwolek and Michal Kepski.
\newblock Human fall detection on embedded platform using depth maps and wireless accelerometer.
\newblock \emph{Computer methods and programs in biomedicine}, 117\penalty0 (3):\penalty0 489--501, 2014.

\bibitem[Le et~al.(2025{\natexlab{a}})Le, Nguyen, Thwal, Qiao, Zhang, and Hong]{fedmekt}
Huy~Q Le, Minh~NH Nguyen, Chu~Myaet Thwal, Yu~Qiao, Chaoning Zhang, and Choong~Seon Hong.
\newblock Fedmekt: Distillation-based embedding knowledge transfer for multimodal federated learning.
\newblock \emph{Neural Networks}, 183:\penalty0 107017, 2025{\natexlab{a}}.

\bibitem[Le et~al.(2025{\natexlab{b}})Le, Thwal, Qiao, Tun, Nguyen, Huh, and Hong]{mfcpl}
Huy~Q Le, Chu~Myaet Thwal, Yu~Qiao, Ye~Lin Tun, Minh~NH Nguyen, Eui-Nam Huh, and Choong~Seon Hong.
\newblock Cross-modal prototype based multimodal federated learning under severely missing modality.
\newblock \emph{Information Fusion}, page 103219, 2025{\natexlab{b}}.

\bibitem[Li et~al.(2021)Li, He, and Song]{moon}
Qinbin Li, Bingsheng He, and Dawn Song.
\newblock Model-contrastive federated learning.
\newblock In \emph{CVPR}, 2021.

\bibitem[Li et~al.(2020)Li, Sahu, Zaheer, Sanjabi, Talwalkar, and Smith]{fedprox}
Tian Li, Anit~Kumar Sahu, Manzil Zaheer, Maziar Sanjabi, Ameet Talwalkar, and Virginia Smith.
\newblock Federated optimization in heterogeneous networks.
\newblock \emph{Proceedings of Machine learning and systems}, 2020.

\bibitem[Lin et~al.(2021)Lin, Lou, Xiong, and Shahabi]{semifed}
Haowen Lin, Jian Lou, Li~Xiong, and Cyrus Shahabi.
\newblock Semifed: Semi-supervised federated learning with consistency and pseudo-labeling.
\newblock \emph{ArXiv}, 2021.

\bibitem[Lin et~al.(2023)Lin, Gao, Gong, Zhang, Zhang, and Li]{mm_survey}
Yi-Ming Lin, Yuan Gao, Mao-Guo Gong, Si-Jia Zhang, Yuan-Qiao Zhang, and Zhi-Yuan Li.
\newblock Federated learning on multimodal data: A comprehensive survey.
\newblock \emph{Machine Intelligence Research}, 20\penalty0 (4):\penalty0 539--553, 2023.

\bibitem[Liu et~al.(2025{\natexlab{a}})Liu, Zeng, Wang, Gao, and Jin]{fedrecon}
Junming Liu, Guosun Zeng, Ding Wang, Yanting Gao, and Yufei Jin.
\newblock Fedrecon: Missing modality reconstruction in distributed heterogeneous environments.
\newblock \emph{arXiv preprint arXiv:2504.09941}, 2025{\natexlab{a}}.

\bibitem[Liu et~al.(2025{\natexlab{b}})Liu, Sun, Liang, Liu, and Xue]{fuels}
Qingxiang Liu, Sheng Sun, Yuxuan Liang, Min Liu, and Jingjing Xue.
\newblock Personalized federated learning for spatio-temporal forecasting: A dual semantic alignment-based contrastive approach.
\newblock In \emph{Proceedings of the AAAI Conference on Artificial Intelligence}, volume~39, pages 12192--12200, 2025{\natexlab{b}}.

\bibitem[Liu et~al.(2024)Liu, Han, Li, and Liu]{semidfl}
Xinyang Liu, Pengchao Han, Xuan Li, and Bo~Liu.
\newblock Semidfl: A semi-supervised paradigm for decentralized federated learning, 2024.

\bibitem[Long et~al.(2020)Long, Tan, Jiang, and Zhang]{finance1}
Guodong Long, Yue Tan, Jing Jiang, and Chengqi Zhang.
\newblock Federated learning for open banking.
\newblock In \emph{Federated learning: privacy and incentive}, pages 240--254. Springer, 2020.

\bibitem[McMahan et~al.(2017)McMahan, Moore, Ramage, Hampson, and Arcas]{fedavg}
Brendan McMahan, Eider Moore, Daniel Ramage, Seth Hampson, and Blaise Aguera~y Arcas.
\newblock {Communication-Efficient Learning of Deep Networks from Decentralized Data}.
\newblock In \emph{Proceedings of International Conference on Artificial Intelligence and Statistics}. PMLR, 2017.

\bibitem[Monfort et~al.(2019)Monfort, Andonian, Zhou, Ramakrishnan, Bargal, Yan, Brown, Fan, Gutfruend, Vondrick, et~al.]{mit}
Mathew Monfort, Alex Andonian, Bolei Zhou, Kandan Ramakrishnan, Sarah~Adel Bargal, Tom Yan, Lisa Brown, Quanfu Fan, Dan Gutfruend, Carl Vondrick, et~al.
\newblock Moments in time dataset: one million videos for event understanding.
\newblock \emph{IEEE Transactions on Pattern Analysis and Machine Intelligence}, pages 1--8, 2019.
\newblock ISSN 0162-8828.
\newblock \doi{10.1109/TPAMI.2019.2901464}.

\bibitem[Nguyen et~al.(2021)Nguyen, Ding, Pathirana, Seneviratne, Li, Niyato, and Poor]{iot1}
Dinh~C Nguyen, Ming Ding, Pubudu~N Pathirana, Aruna Seneviratne, Jun Li, Dusit Niyato, and H~Vincent Poor.
\newblock Federated learning for industrial internet of things in future industries.
\newblock \emph{IEEE Wireless Communications}, 28\penalty0 (6):\penalty0 192--199, 2021.

\bibitem[Nguyen et~al.(2022)Nguyen, Pham, Pathirana, Ding, Seneviratne, Lin, Dobre, and Hwang]{health3}
Dinh~C Nguyen, Quoc-Viet Pham, Pubudu~N Pathirana, Ming Ding, Aruna Seneviratne, Zihuai Lin, Octavia Dobre, and Won-Joo Hwang.
\newblock Federated learning for smart healthcare: A survey.
\newblock \emph{ACM Computing Surveys (Csur)}, 55\penalty0 (3):\penalty0 1--37, 2022.

\bibitem[Nguyen et~al.(2024)Nguyen, Nguyen, Pham, Hoang, Le~Nguyen, and Huynh]{fedmac}
Manh~Duong Nguyen, Trung~Thanh Nguyen, Huy~Hieu Pham, Trong~Nghia Hoang, Phi Le~Nguyen, and Thanh~Trung Huynh.
\newblock Fedmac: Tackling partial-modality missing in federated learning with cross-modal aggregation and contrastive regularization.
\newblock In \emph{2024 22nd International Symposium on Network Computing and Applications (NCA)}, pages 278--285. IEEE, 2024.

\bibitem[Poria et~al.(2018)Poria, Hazarika, Majumder, Naik, Cambria, and Mihalcea]{meld}
Soujanya Poria, Devamanyu Hazarika, Navonil Majumder, Gautam Naik, Erik Cambria, and Rada Mihalcea.
\newblock Meld: A multimodal multi-party dataset for emotion recognition in conversations.
\newblock \emph{arXiv preprint arXiv:1810.02508}, 2018.

\bibitem[Qu et~al.(2022)Qu, Zhou, Liang, Xia, Wang, Adeli, Fei-Fei, and Rubin]{rethink_data}
Liangqiong Qu, Yuyin Zhou, Paul~Pu Liang, Yingda Xia, Feifei Wang, Ehsan Adeli, Li~Fei-Fei, and Daniel Rubin.
\newblock Rethinking architecture design for tackling data heterogeneity in federated learning.
\newblock In \emph{cvpr}, 2022.

\bibitem[Reddi et~al.(2020)Reddi, Charles, Zaheer, Garrett, Rush, Konecny, Kumar, and McMahan]{fedopt}
Sashank Reddi, Zachary Charles, Manzil Zaheer, Zachary Garrett, Keith Rush, Jakub Konecny, Sanjiv Kumar, and H~Brendan McMahan.
\newblock Adaptive federated optimization.
\newblock \emph{arXiv preprint arXiv:2003.00295}, 2020.

\bibitem[Saunshi et~al.(2019)Saunshi, Plevrakis, Arora, Khodak, and Khandeparkar]{ae2}
Nikunj Saunshi, Orestis Plevrakis, Sanjeev Arora, Mikhail Khodak, and Hrishikesh Khandeparkar.
\newblock A theoretical analysis of contrastive unsupervised representation learning.
\newblock In \emph{International Conference on Machine Learning}, pages 5628--5637. PMLR, 2019.

\bibitem[Song et~al.(2024)Song, Yang, Zhang, Fu, Xu, and King]{semi_survey}
Zixing Song, Xiangli Yang, Yifei Zhang, Xinyu Fu, Zenglin Xu, and Irwin King.
\newblock A systematic survey on federated semi-supervised learning.
\newblock In \emph{IJCAI}, 2024.

\bibitem[Sun et~al.(2020)Sun, Yu, Song, Liu, Yang, and Zhou]{mobilebert}
Zhiqing Sun, Hongkun Yu, Xiaodan Song, Renjie Liu, Yiming Yang, and Denny Zhou.
\newblock Mobilebert: a compact task-agnostic bert for resource-limited devices.
\newblock \emph{arXiv preprint arXiv:2004.02984}, 2020.

\bibitem[Tashakori et~al.(2023)Tashakori, Zhang, Wang, and Servati]{semipfl}
Arvin Tashakori, Wenwen Zhang, Z.~Wang, and Peyman Servati.
\newblock Semipfl: Personalized semi-supervised federated learning framework for edge intelligence.
\newblock \emph{IEEE Internet of Things Journal}, 2023.

\bibitem[Teo et~al.(2024)Teo, Jin, Liu, Li, Miao, Zhang, Ng, Tan, Lee, Chua, et~al.]{health1}
Zhen~Ling Teo, Liyuan Jin, Nan Liu, Siqi Li, Di~Miao, Xiaoman Zhang, Wei~Yan Ng, Ting~Fang Tan, Deborah~Meixuan Lee, Kai~Jie Chua, et~al.
\newblock Federated machine learning in healthcare: A systematic review on clinical applications and technical architecture.
\newblock \emph{Cell Reports Medicine}, 5\penalty0 (2), 2024.

\bibitem[Vincent et~al.(2010)Vincent, Larochelle, Lajoie, Bengio, Manzagol, and Bottou]{ae1}
Pascal Vincent, Hugo Larochelle, Isabelle Lajoie, Yoshua Bengio, Pierre-Antoine Manzagol, and Leon Bottou.
\newblock Stacked denoising autoencoders: Learning useful representations in a deep network with a local denoising criterion.
\newblock \emph{Journal of machine learning research}, 11\penalty0 (12), 2010.

\bibitem[Wen et~al.(2023)Wen, Zhang, Lan, Cui, Cai, and Zhang]{fed_survey}
Jie Wen, Zhixia Zhang, Yang Lan, Zhihua Cui, Jianghui Cai, and Wensheng Zhang.
\newblock A survey on federated learning: challenges and applications.
\newblock \emph{International Journal of Machine Learning and Cybernetics}, 14\penalty0 (2):\penalty0 513--535, 2023.

\bibitem[Yan and Guo(2024)]{lufl}
Hao Yan and Yuhong Guo.
\newblock Lightweight unsupervised federated learning with pretrained vision language model, 2024.

\bibitem[Yang et~al.(2024)Yang, Yu, McKeown, Wang, et~al.]{health2}
Tian Yang, Xinhui Yu, Martin~J McKeown, Z~Jane Wang, et~al.
\newblock When federated learning meets medical image analysis: A systematic review with challenges and solutions.
\newblock \emph{APSIPA Transactions on Signal and Information Processing}, 13\penalty0 (1), 2024.

\bibitem[Ye et~al.(2023)Ye, Fang, Du, Yuen, and Tao]{heter_fl}
Mang Ye, Xiuwen Fang, Bo~Du, Pong~C Yuen, and Dacheng Tao.
\newblock Heterogeneous federated learning: State-of-the-art and research challenges.
\newblock \emph{ACM Computing Surveys}, 56\penalty0 (3):\penalty0 1--44, 2023.

\bibitem[Yu et~al.(2023)Yu, Liu, Wang, Xu, and Liu]{creamfl}
Qiying Yu, Yang Liu, Yimu Wang, Ke~Xu, and Jingjing Liu.
\newblock Multimodal federated learning via contrastive representation ensemble.
\newblock \emph{ICLR}, 2023.

\bibitem[Zhao et~al.(2020)Zhao, Liu, Li, Barnaghi, and Haddadi]{ssfl4ar}
Yuchen Zhao, Hanyang Liu, Honglin Li, Payam~M. Barnaghi, and Hamed Haddadi.
\newblock Semi-supervised federated learning for activity recognition.
\newblock \emph{ArXiv}, abs/2011.00851, 2020.
\newblock URL \url{https://api.semanticscholar.org/CorpusID:226226498}.

\bibitem[Zhao et~al.(2022)Zhao, Barnaghi, and Haddadi]{mmfed}
Yuchen Zhao, Payam Barnaghi, and Hamed Haddadi.
\newblock Multimodal federated learning on iot data.
\newblock In \emph{IEEE/ACM International Conference on Internet-of-Things Design and Implementation (IoTDI)}, 2022.

\bibitem[Zheng et~al.(2024)Zheng, Liu, Liu, and Tan]{fedtccl}
Haowen Zheng, Hui Liu, Zhenyu Liu, and Jianrong Tan.
\newblock Federated temporal-context contrastive learning for fault diagnosis using multiple datasets with insufficient labels.
\newblock \emph{Advanced Engineering Informatics}, 60:\penalty0 102432, 2024.

\end{thebibliography}
\end{document}